\documentclass[superscriptaddress, twocolumn, amsmath,amssymb,aps,prl]{revtex4-2}

\usepackage{graphicx}
\usepackage{dcolumn}
\usepackage{bm}
\usepackage{caption}
\usepackage{subcaption}
\usepackage{float}
\usepackage[mathlines]{lineno}
\usepackage{multirow}

\begin{document}


\title{Direct constraints on ultra-light boson mass from searches for continuous gravitational waves}

\author{C. Palomba}
\affiliation{INFN, Sezione di Roma, I-00185 Roma, Italy}%
\author{S. D'Antonio}%
\affiliation{INFN, Sezione di Roma Tor Vergata, I-00133 Roma , Italy}
\author{P. Astone}
\affiliation{INFN, Sezione di Roma, I-00185 Roma, Italy}%
\author{S. Frasca}
\affiliation{University of Rome La Sapienza, I-00185 Roma, Italy}%
\affiliation{INFN, Sezione di Roma, I-00185 Roma, Italy}%
\author{G. Intini}
\affiliation{University of Rome La Sapienza, I-00185 Roma, Italy}%
\affiliation{INFN, Sezione di Roma, I-00185 Roma, Italy}%
\author{I. La Rosa}
\affiliation{Laboratoire dÕAnnecy-le-Vieux de Physique des Particules (LAPP), Universit?e Savoie Mont Blanc, CNRS/IN2P3, F-74941 Annecy, France}
\author{P. Leaci}
\affiliation{University of Rome La Sapienza, I-00185 Roma, Italy}%
\affiliation{INFN, Sezione di Roma, I-00185 Roma, Italy}%
\author{S. Mastrogiovanni}
\affiliation{PC,  AstroParticule  et  Cosmologie,  Universit?e  Paris  Diderot, CNRS/IN2P3,  CEA/Irfu,  Observatoire  de  Paris, Sorbonne  Paris  Cit?e,  F-75205  Paris  Cedex  13,  France}
\author{A. L. Miller}
\affiliation{University of Rome La Sapienza, I-00185 Roma, Italy}%
\affiliation{INFN, Sezione di Roma, I-00185 Roma, Italy}%
\affiliation{University of Florida, Gainseville, Florida 32611, USA}
\author{F. Muciaccia}
\affiliation{University of Rome La Sapienza, I-00185 Roma, Italy}%
\author{O. J. Piccinni}
\affiliation{University of Rome La Sapienza, I-00185 Roma, Italy}%
\affiliation{INFN, Sezione di Roma, I-00185 Roma, Italy}%
\author{L. Rei}
\affiliation{INFN, Sezione di Genova, I-16146, Italy}
\author{F. Simula}
\affiliation{INFN, Sezione di Roma, I-00185 Roma, Italy}%

\date{\today}

\begin{abstract}
\textit{Superradiance} can trigger the formation of an ultra-light boson cloud around a spinning black hole. Once formed, the boson cloud is expected to emit a nearly periodic, long-duration, gravitational-wave signal. For boson masses in the range $(10^{-13}-10^{-11})$ eV, and stellar mass black holes, such signals are potentially detectable by gravitational wave detectors, like Advanced LIGO and Virgo. 
In this {\it Letter} we present full band upper limits for a generic all-sky search for periodic gravitational waves in LIGO O2 data, and use them to derive - for the first time - direct constraints on the ultra-light scalar boson field mass.
\end{abstract}

\maketitle


\paragraph{Introduction.---}
Ultra-light bosons with masses $m_\text{b} \ll$ 1 eV, including e.g. axions or dark photons, are predicted in theories beyond the Standard Model and could also be a component of dark matter, see e.g. \cite{ref:cardo2018}. Light boson fields  can scatter on spinning black holes, and the scattered field amplitude be amplified due to the {\it superradiance} process, see \cite{ref:brito2015} for a review. This amplification, which takes place at the expense of the black hole angular momentum, creates a classical boson condensates (a ``cloud'') around the black hole itself. 
The process stops after a time $\tau_\text{inst}$, when the real part of the boson field angular frequency is equal to the black hole horizon angular frequency. Once equilibrium has been reached, the boson cloud mass is dissipated by emitting a nearly monochromatic gravitational wave signal \cite{ref:arva2015,ref:brito2017,ref:arva2017,ref:brito2017b}, with a frequency equal to two times that of the field and given, for the dominant mode of a scalar field, by \cite{ref:brito2017}
\begin{equation}
\frac{f_\text{gw}}{1~\text{Hz}}\simeq 483 \left(\frac{m_\text{b}}{10^{-12}~\text{eV}}\right)
\left[1-\frac{6}{25}\left(\frac{M_\text{bh}}{10M_{\odot}}\frac{m_\text{b}}{10^{-12}~\text{eV}}\right)^2\right].
\label{eq:fgw}
\end{equation}
This means that if bosons have mass in the range $(10^{-13}-10^{-11})~\text{eV}$, then the signal emitted by a cloud around a stellar mass black hole, with mass $(10-100)~M_{\odot}$, is in the sensitivity band of Advanced LIGO \cite{ref:ligo} and Virgo \cite{ref:virgo} detectors. The emission takes place over a timescale $\tau_{\text{GW}}$ typically much longer than the detector observation times, and has an amplitude decaying in time as $\left(1+t/\tau_\text{GW}\right)^{-1}$ \cite{ref:brito2017}. 

The search for these periodic signals can be done using data analysis techniques similar to those used for the search of persistent signals emitted, for instance, by asymmetric spinning neutron stars, which are a typical target - although still not detected - of the LIGO and Virgo detectors. Specifically, for a given source direction, it is necessary to take into account the Doppler effect due to the Earth's rotation and revolution around the Sun (plus a further correction if the source is in a binary system \cite{ref:leaci}) and the long-term frequency variation (spin-down or spin-up). 
Moreover, the sidereal effect due to detector beam pattern functions, which causes an amplitude and phase modulation, and smaller relativistic effects, namely the Einstein delay and the Shapiro delay, may need to be considered. In the absence of un-modeled features, the expected signal becomes monochromatic, after the various effects described above have been properly taken into account. See \cite{ref:keith2017}, \cite{ref:palo2017} for recent reviews on continuous wave searches and \cite{ref:bayes,ref:targfstat,ref:fivevect,ref:nb_method1,ref:nb_method2,ref:fstat,ref:skyhough,ref:freqhough,ref:powerflux,ref:einstein,ref:direct,ref:whelan,ref:goetz,ref:messenger,ref:sourova} for a more detailed description of various search methods. 

In particular, \textit{all-sky} searches for periodic sources with no electromagnetic counterpart generally cover a wide portion of the search parameter space, which consists of the source sky position, signal frequency and frequency derivative. The typical procedure that we use is a semi-coherent approach where the data time series is divided in several chunks of a given duration $T_\text{FFT}$, which are then properly combined dropping the requirement of phase consistency among the various data segments \cite{ref:fstat,ref:skyhough,ref:freqhough,ref:powerflux,ref:einstein}. This makes the search feasible, but still computationally costly, at the price of a sensitivity loss \cite{ref:freqhough} with respect to full coherent searches, see e.g. \cite{ref:targ_o2}. The duration $T_\text{FFT}$ is typically chosen such that the Doppler modulation is confined within a single frequency bin (i.e. $1/T_\text{FFT}$),  or longer if a partial correction of the Doppler effect is done before. In such methods the sensitivity to monochromatic signals, i.e. the minimum detectable amplitude, goes as $\left(T_\text{obs}\times T_\text{FFT}\right)^{-1/4}$, where $T_\text{obs}$ is the total observation time. 
They have been applied to several LIGO and Virgo runs [see \cite{ref:allsky_o1,ref:einstein_o1,ref:allsky_o1_full,ref:power_o1,ref:allsky_o2} for the most recent results to date]. 
For instance, the all-sky hierarchical pipeline based on the \textit{FrequencyHough} transform \cite{ref:freqhough} used $T_\text{FFT}=8192, ~4096,~ 2048,~ 1024$ second for the frequency ranges $(10-128)$ Hz, $(128-512)$ Hz, $(512-1024)$ Hz and $(1024-2048)$ Hz respectively, to analyze LIGO O2 data, which covered about 9 months from November 2016 to August 2017 \cite{ref:allsky_o2}. Virgo data have not been used due to the shorter data taking time ($\sim$1 month) and worse sensitivity.  As no significant candidate has been found, 95\% confidence level (C. L.) upper limits on the signal strain amplitude have been computed.

In this paper we map these upper limits, which are the best obtained so far in any all-sky search for periodic gravitational waves, in exclusion regions in the plane defined by the mass of the {\it scalar} boson field and the mass of the black hole, assuming the emitted signals are nearly monochromatic. Such kinds of constraints have been recently studied mainly in the context of future third-generation detectors, specifically the Cosmic Explorer \cite{ref:cosmic}, and considering boson clouds around post-merger black holes placed at extra-galactic distances \cite{ref:ghosh2019}. On the other hand, we present here constraints computed - for the first time - using actual results from the latest O2 all-sky searches for periodic gravitational waves. We notice that an interesting, although less comprehensive discussion on the interpretation of LIGO O1 data all-sky search results (over the frequency range 20-200 Hz) in terms of distance reach to {\it vector} boson condensates has been briefly touched upon in \cite{ref:dergachev2019}. 



\paragraph{O2 all-sky search upper limits.---}
The most sensitive all-sky searches for periodic gravitational wave signals have been described in \cite{ref:allsky_o2}, where results by three independent pipelines have been presented. In particular, the \textit{FrequencyHough} pipeline covered a parameter space shown in the first row of Tab. \ref{tab:table1}. 
\begin{table}[h!]
    \caption{Parameter space covered by the initial FrequencyHough search \cite{ref:allsky_o2} and by the new extended search.}
    \label{tab:table1}
    \begin{ruledtabular}
    \begin{tabular}{lc|c|c|}
    search & frequency range [Hz] & Spin-down range [Hz/s] \\ \hline
    \multirow{2}{*}{initial} & $(10,~512)$ &  $-10^{-8},~ 2\times 10^{-9}$ \\ 
    			 & $(512,~1024)$ &  $-2\times 10^{-9},~ 2\times 10^{-9}$ \\  \hline
    extended & $(10,~2048)$ &  $-10^{-8},~ 2\times 10^{-9}$ \\ 
    \end{tabular}
 \end{ruledtabular}
\end{table} 
Population-based upper limits have been obtained by injecting in each 1 Hz band several sets of signals, each with a fixed amplitude and random parameters, and finding the signal amplitude such that 95\% of the injected signals were recovered by the pipeline with a Critical Ratio (defined as $CR=\left(x-\mu_x\right)/\sigma_x$, where $x$ the number count on the FrequencyHough map, $\mu_x$ and $\sigma_x$ are the mean value and the standard deviation) $\ge$ than that of the loudest candidate found in the real search.   

Here we present the latest upper limits obtained with the \textit{FrequencyHough} pipeline, extending the search parameter space both in frequency and spin-down, see the second row in Tab. \ref{tab:table1}. 
For computational efficiency reasons, these extended upper limits have been obtained with a faster approximate procedure. The first step consists in using the previous \textit{FrequencyHough} upper limits $h^\text{old}_{0,\text{ul}}(f)$, computed in each 1-Hz band at frequency smaller than $1024$ Hz with the full injection procedure, and described in \cite{ref:allsky_o2}, to find the ratios $K(f)$ with a sensitivity estimation $h_\text{sens}(f)$, based on Eq. (67) of \cite{ref:freqhough}. The latter has been computed using the data average power spectrum and a fixed reference critical ratio value $CR_\text{ref}=3$ in the factor $\sqrt{CR_\text{ref}-\sqrt{2}\text{erfc}^{-1}(2\Gamma)}$, where $\Gamma=0.95$ is the chosen C.L. Next, we have evaluated the median value of the $K(f)$, $\bar{K}_\text{low}$ (where {\it low} stands for low frequency), which has been used to re-calculate the upper limits in the range (10-1024) Hz as 
\begin{equation}
h_{0,\text{ul}}(f) \approx \bar{K}_\text{low}\times  h_\text{sens}(f)\frac{\sqrt{CR_\text{max}-\sqrt{2}\text{erfc}^{-1}(2\Gamma)}}{\sqrt{CR_\text{ref}-\sqrt{2}\text{erfc}^{-1}(2\Gamma)}},
\label{eq:approxul}
\end{equation} 
where $CR_\text{max}$ is the actual candidate maximum critical ratio value in the 1-Hz band at frequency $f$. The 1-$\sigma$ percentile of the quantity $|\frac{h^\text{old}_{0,\text{ul}}-{h_{0,\text{ul}}}}{h_{0,\text{ul}}}|$, $\sigma_\text{ul,low}\simeq 0.13$, is taken as a measure of the typical relative ``error'' of the estimation (there are, of course, cases in which the actual difference is larger, especially for strongly disturbed frequency bands). Given this result, which demonstrates the fast computation is reliable, we extend the upper limit to the range (1024-2048) Hz. First, we inject signals in 30 different bands and compute the ``exact'' upper limit at the corresponding frequencies. Then the approximate procedure is followed again, by finding the ratios of the upper limits to the sensitivity estimation and the corresponding median value $\bar{K}_\text{high}$  (where {\it high} stands for high frequency). By means of Eq. \ref{eq:approxul}, with the replacement $\bar{K}_\text{low} \rightarrow \bar{K}_\text{high}$, the upper limits for all the 1-Hz bands in the range (1024-2048) Hz are computed. Although injections have been done in a much smaller number of 1-Hz bands, which allowed us to significantly reduce the computational time, the accuracy for the range (1024-2048) Hz is $\sigma_\text{ul,high}\simeq \sigma_\text{ul,low}$.
 
\begin{figure}[h]
\includegraphics[width=0.5\textwidth]{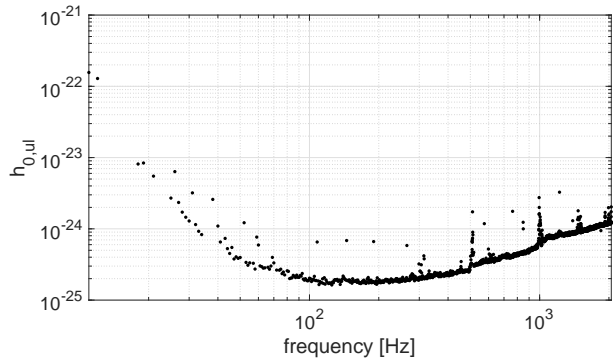}
\caption{\label{fig:ul_all}Full 95\% O2 upper limits on the signal strain amplitude obtained with the \textit{FrequencyHough} pipeline. Each dot represents the upper limit in a 1-Hz band between 10 and 2048 Hz, computed with the fast procedure described in the text.}
\end{figure}

Upper limits over the whole frequency range (10-2048) Hz are shown in Fig. \ref{fig:ul_all}. 
In the range 1024-1500 Hz our results are slightly better, within $\sim 20\%$, w.r.t. those obtained by the \textit{SkyHough} pipeline \cite{ref:allsky_o2}, which use in this frequency range similar values of $T_\text{FFT}$. Above 1500 Hz the results we find are a factor $\sim$ 2.5 better than those obtained by the \textit{Time Domain F-statistic} pipeline \cite{ref:allsky_o2}. 

\begin{figure*}[h]
\begin{subfigure}{8cm}
\centering
\includegraphics[width=8.5cm]{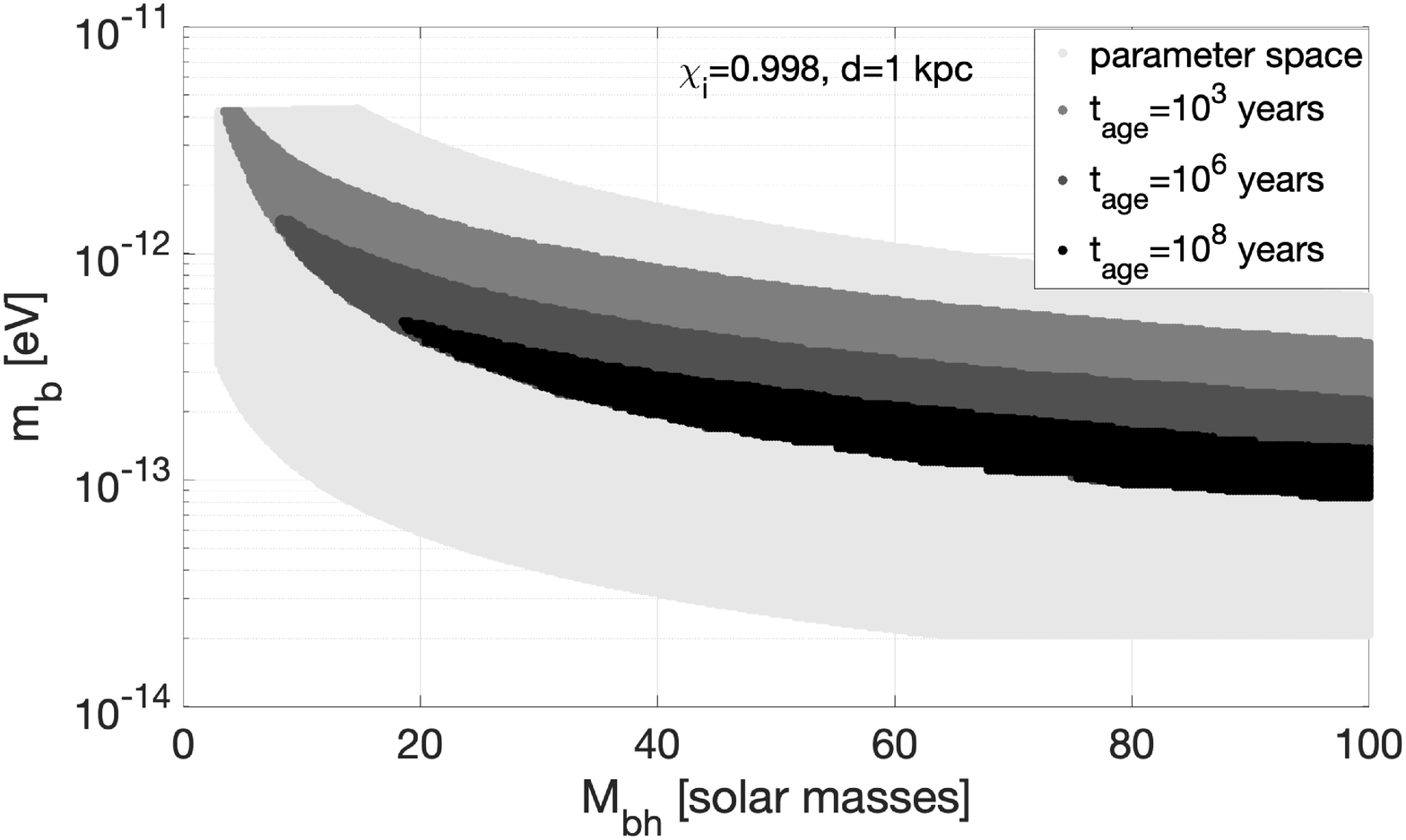}
\end{subfigure}
\begin{subfigure}{8cm}
\centering
\includegraphics[width=8.5cm]{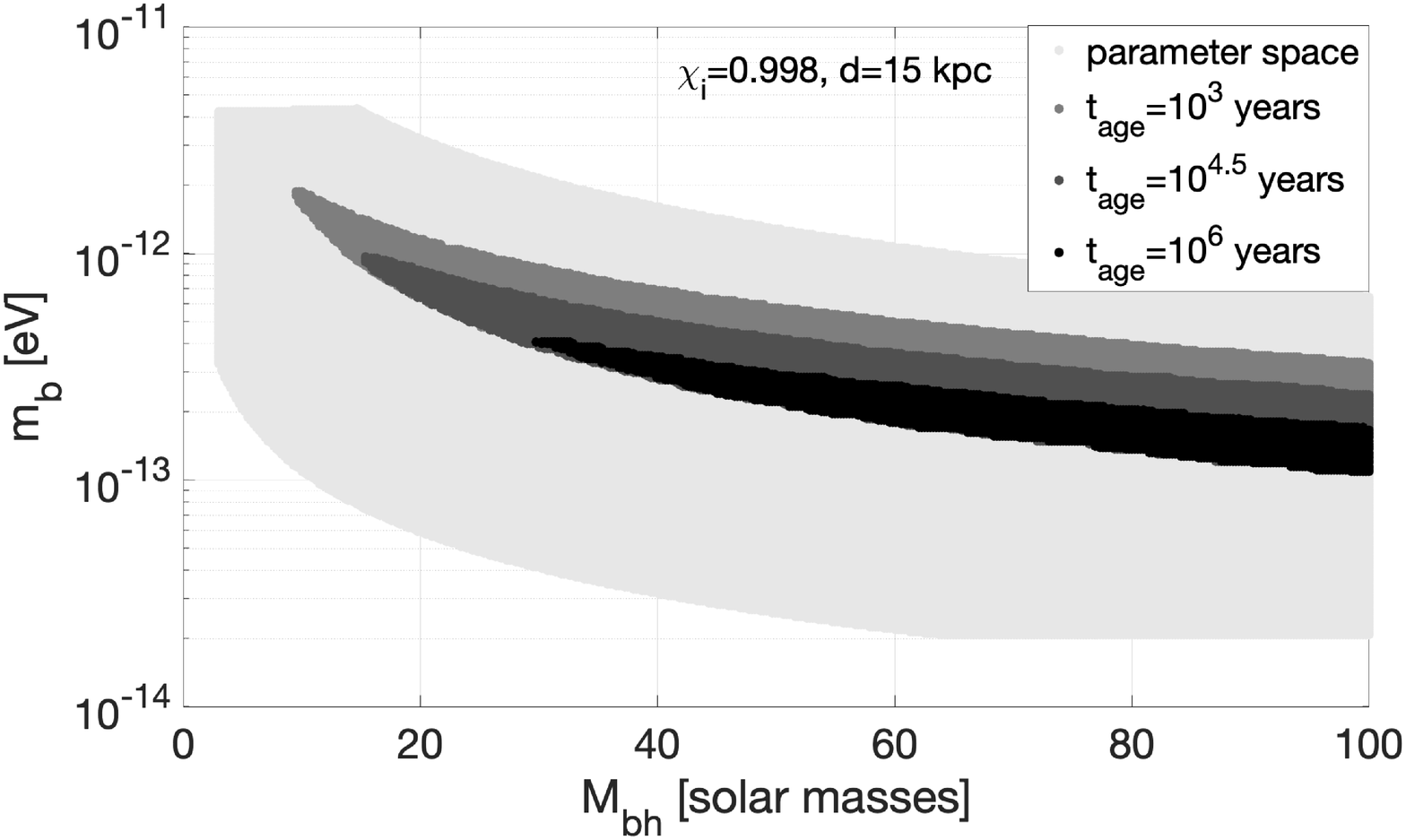}
\end{subfigure}
\caption{95\% C.L. exclusion regions in the plane $m_\text{b} - M_\text{bh}$ assuming a maximum distance $d=1$ kpc (left plot) and $d=15$ kpc (right plot), a black hole initial a-dimensional spin $\chi_i=0.998$, and three possible values for $t_\text{age}$: $10^3$, $10^6$, $10^8$ years (left plot) and $10^3$, $10^{4.5}$, $10^6$ years (right plot). The larger light gray area is the accessible parameter space. As expected, the extension of the excluded region decreases for increasing $t_\text{age}$ (corresponding to darker color). }
\label{fig:exclusion_chi0998}
\end{figure*}

\begin{figure*}[h]
\begin{subfigure}{8cm}
\includegraphics[width=8.5cm]{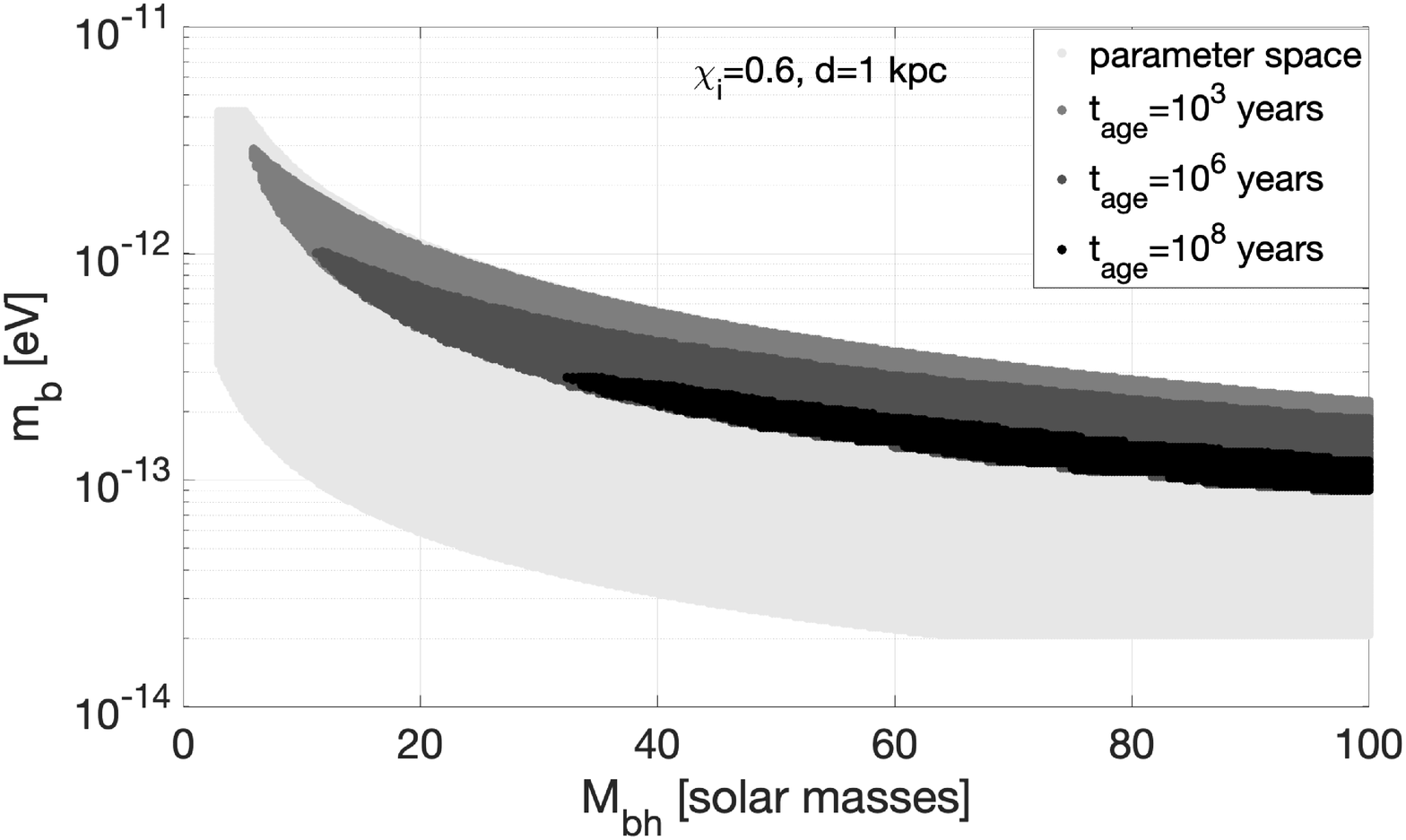}
\label{fig:exclusion_15kpc_chi099}
\end{subfigure}
\begin{subfigure}{8cm}
\includegraphics[width=8.5cm]{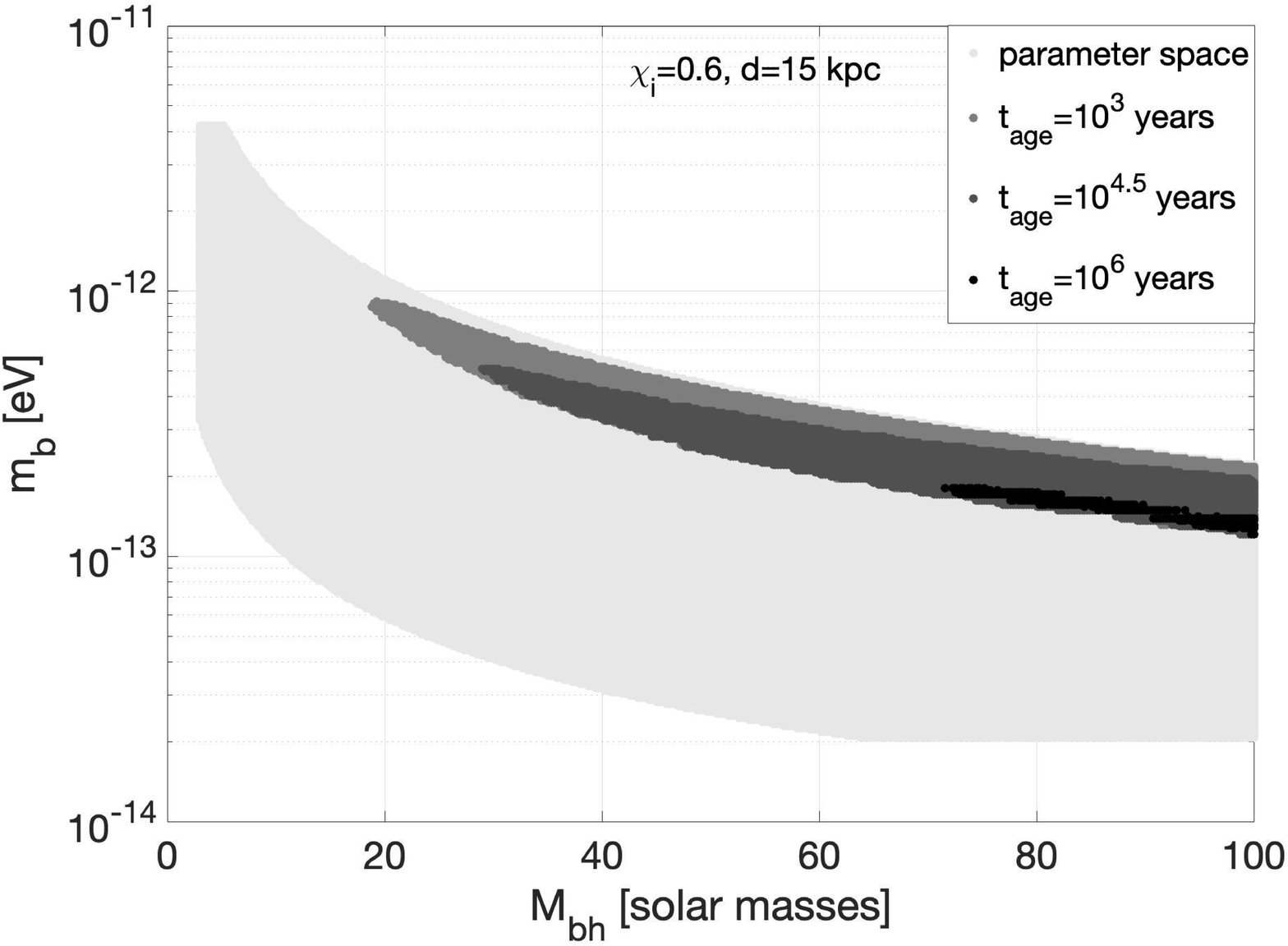}
\label{fig:exclusion_1kpc_chi06}
\end{subfigure}
\caption{95\% C.L. exclusion regions in the plane $m_\text{b} - M_\text{bh}$ assuming a maximum distance $d=1$ kpc (left plot) and $d=15$ kpc (right plot), a black hole initial a-dimensional spin $\chi_i=0.6$, and three possible values for $t_\text{age}$: $10^3,~10^6,~10^8$ years (left plot) and $10^3$, $10^{4.5}$, $10^6$ years (right plot). Color code as in previous figure. As expected, the excluded region decreases for increasing $t_\text{age}$.}
\label{fig:exclusion_chi06}
\end{figure*}

\paragraph{Exclusion regions.---}
We consider a range $M_\text{bh}\in [3, 100~M_{\odot}]$ for the black hole mass  and $m_\text{b}\in [10^{-14}, 10^{-11}]$ eV for the bosons. For each pair $(m_\text{b}, ~M_\text{bh})$, we first verify if certain conditions are met, assuming the emission is dominated by the fundamental mode of a scalar field. This allows to determine the parameter space potentially accessible by the search. First, the signal frequency computed by Eq. \ref{eq:fgw} must be within the search range, i.e $f_\text{gw}\in [10, 2048]$ Hz. At the same time, it must be also smaller than  two times the black hole initial spin frequency, which is the condition for the {\it superradiance} process to happen: $f_\text{gw}<\Omega_H/\pi=c^3\mathcal{R}(\chi_i)/(2\pi GM_\text{bh})$, where $\chi_i\in [0,1)$ is the black hole initial adimensional spin and $\mathcal{R}(\chi_i)=\chi_i/(1+\sqrt{1-\chi_i^2})$ \cite{ref:isi2019}. Second, we impose that the {\it superradiance} time scale  $\tau_\text{inst}$ is smaller than the Hubble time \cite{ref:danto2018}
\begin{equation}
\left(\frac{m_\text{b}}{10^{-13}~eV}\right)\ge 0.752  \left(\frac{M_\text{bh}}{10M_{\odot}}\right)^{-8/9}\chi^{-1/9}_f,
\label{eq:tauincond}
\end{equation}
where $\chi_f$ is the black hole spin at the end of the \textit{superradiance} phase. Third,
we require that it is also much shorter (at least ten times) than the gravitational wave emission time scale $\tau_\text{GW}$ as, otherwise, the \textit{superradiance} process would not reach saturation and the gravitational radiation emission would be significantly reduced \cite{ref:brito2017}.
The corresponding parameter space is shown as a light gray region in Figures \ref{fig:exclusion_chi0998} (for $\chi_i=0.998$) and  \ref{fig:exclusion_chi06} (for $\chi_i=0.6$). 

We assign now a value to the black hole initial spin $\chi_i$, to the time since the beginning of the emission $t_\text{age}$, and to the distance $d$. Hence, from \cite{ref:brito2017} we compute - for each pair $(m_\text{b}, ~M_\text{bh})$ - the signal amplitude $h_0$ at the detector (averaged over sky position) as 
\begin{eqnarray}
h_0(t_\text{age})\simeq 1.15\times 10^{-21}\left(\frac{\dot{\tilde{E}}}{10^{-12}}\right)^{1/2}\left(\frac{M_S}{M_\text{bh}}\right) \nonumber \\
 \times \left(\frac{m_\text{b} }{5\times 10^{-13}\text{eV}}\right)^{-1} \left(\frac{d}{10\text{kpc}}\right)^{-1}\left(1+\frac{t_\text{age}}{\tau_\text{GW}}\right)^{-1}.
\label{eq:ampl}
\end{eqnarray}
For the rescaled gravitational wave luminosity $\dot{\tilde{E}}=\dot{E}\times\left(M_\text{bh}/M_S\right)^2$, where $M_S(M_\text{bh}, m_\text{b}, \chi_i)$ is the cloud mass at the saturation, we use a 6-th order polynomial fit to the numerical result found by \cite{ref:brito2017}. Such fit is accurate to better than $1$\% over the range $0.1\le M_\text{bh}\times \mu \le 0.5$, where $\mu=\frac{G}{\hbar c^3}m_\text{b}$, and better than 10\% down to $M_\text{bh}\times \mu \simeq 0.0067$, which is the lowest value for the parameter space we are considering. 
Finally, the amplitude is compared to the upper limit $h_\text{0,ul}$ at the signal frequency $f_\text{gw}$. If $h_0(t_\text{age},\chi_i,d)> h_\text{0,ul}$ we exploit the non-detection in the O2 all-sky search to conclude that a boson cloud - black hole system with those masses, and emitting since a time $t_\text{age}$, cannot be present within a distance $d$. In fact, to be conservative we increase the upper limit values by a factor 1.13 to take into account the upper limit uncertainty discussed previously. Note that the upper limits have been obtained in a search covering a maximum spin-up much larger than the predicted spin-up of the  signal emitted by the boson cloud, see \cite{ref:arva2015}. Tab. \ref{tab:table2} describes the parameter values used for the result plots.
\begin{table}[h!]
    \caption{Values of source distance $d$, initial adimensional black hole spin $\chi_i$ and age $t_\text{age}$ used to produce the result plots. See text for more details.}
    \label{tab:table2}
    \begin{tabular}{c|c|c}
    \toprule
    distance [kpc] & initial BH spin & age [yr] \\ \hline
    1 & 0.6, 0.998 &  $10^{3}, 10^{6}, 10^{8}$ \\ 
    15	 & 0.6, 0.998 &  $10^{3}, 10^{4.5}, 10^{6}$ \\
    \hline  
    \end{tabular}
\end{table} 
These values cover different scenarios going from optimistic (near source, high initial spin, very young age) to more realistic (source at galactic distance, moderate spin, middle to old age).
 Left plot in Figure \ref{fig:exclusion_chi0998} shows results assuming a maximum distance $d=1$ kpc, an initial black hole a-dimensional spin $\chi_i=0.998$, and three possible values for $t_\text{age}$: $10^3,~10^6,~10^8$ years, while the right plot is for $d=15$ kpc and $t_\text{age}$: $10^3,~10^{4.5},~10^6$ years.
Most of the excluded parameter space lies in the range of boson masses between $\sim 10^{-13}$ eV and $\sim 10^{-12}$ eV and - as expected - is smaller for older systems (bigger $t_\text{age}$). 

As a matter of fact, the number and mass distribution of black holes in the Milky Way play a relevant role in establishing actual exclusion regions. Considering a supernova (of type II) rate of 2-3 per century in the Milky way \cite{ref:li2011}, using the Kroupa initial mass function \cite{ref:kroupa} and assuming that progenitor stars with masses larger than about $30~M_\odot$ collapse to black hole, the current expected galactic black hole formation rate is about two per thousand years. Roughly 90\% of them is expected to have mass smaller than about 30 $M_\odot$ and about 1\% mass above $\sim 50~M_\odot$  \cite{ref:elbert2018}. At distances from us smaller than  $d=1$ kpc, the formation rate is $\sim 2\times 10^{-5}$ per year \cite{ref:soren2017}, so we do not expect any black hole with age less or equal to $10^3$ years, hence the exclusion region shown in Fig. \ref{fig:exclusion_chi0998}-left for $t_\text{age}=10^3$ years indeed is not very significant. A few tens black holes with age up to $10^6$ years could be present within the same distance, and only a few of them should have masses above $20-30~M_\odot$, so that boson masses roughly in the range $(1-10)\times 10^{-13}$ eV can be marginally excluded, assuming highly spinning black holes and that the \textit{superradiance} process actually takes place for \textit{every} nearby black hole. 

By considering $d=15$ kpc we have - on average - smaller gravitational wave signals, but a significantly higher number of black holes. Roughly speaking, $\sim 2$ black holes with $t_\text{age} \le 10^3$ years, $\sim 100$ with $t_\text{age} \le 10^{4.5}$ years and $\sim 2\times 10^3$ black holes with $t_\text{age} \le 10^6$ years, should exist. In the latter two cases, we may estimate that $\sim 10$, for $t_\text{age} \le 10^{4.5}$ years, and $\sim 100$ for $t_\text{age} \le 10^6$ years, have masses above 30 $M_\odot$. This means that the range of boson masses defined by the darker color region in Fig. \ref{fig:exclusion_chi0998} (right panel), extending from  $1.1\times 10^{-13}$ eV to about $4\times 10^{-13}$ eV can be excluded with rather strong confidence, even under the hypothesis that only a few percent of the whole galactic black hole population undergoes the \textit{superradiance} process. Assuming most of the galactic black hole population is subject to the \textit{superradiance}, the excluded region can be extended up to about $9\times 10^{-13}$ eV. We have not tried to find the exact distance which would maximise the chance of detection, by taking into account the expected signal amplitude and the black hole formation rate, as this would require a full population simulation (with several uncertain quantities) which is outside the scope of this paper. We expect, however, that going outside the Milky Way does not improve detection probability much (hence the choice of using $d=15$ kpc) as the supernova rate increases only slightly up to the Virgo Cluster. 

In the case $\chi_i=0.6$ the parameter space shrinks as (i) the instability timescale is longer, (ii) the region satisfying the \textit{superradiant} condition, $\pi f_\text{gw}<\Omega_\text{H}$  is smaller. 
Results for this case are shown in Fig. \ref{fig:exclusion_chi06}, again assuming a maximum distance $d=1$ kpc (left plot) and $d=15$ kpc (right plot). By considering the galactic black hole formation rate  and their expected mass distribution, conclusions similar to the previous ones can be drawn. In particular, the most robust prediction we can make is that the light color region shown in Fig. \ref{fig:exclusion_chi06} (right panel), covering the boson mass range $(1.2-1.8)\times 10^{-13}$ eV, can be rather firmly excluded. The excluded range extends up to $\sim 5 \times 10^{-13}$ eV (corresponding to $t_\text{age} \le 10^{4.5}$ years) if most of the galactic black population is subject to the \textit{superradiance}.

Overall, O2 data are such that, even in favorable cases, i.e. high initial black hole spin and newborn clouds (with age of $\sim$ ten years), we cannot be sensitive to signals coming from distances larger than $\sim 5$ Mpc. 
This can be shown to be in agreement with the estimations provided in \cite{ref:brito2017} and \cite{ref:isi2019}, after taking into account the difference in $T_\text{FFT}$, detector sensitivity, observation time among the two cases, and the fact that we have used real data results instead of theoretical sensitivity estimations.


\paragraph{Discussion.---}
Two major accomplishments are described in this {\it Letter}.  First, we have extended from 1024 Hz up to 2048 Hz the upper limits obtained by the \textit{FrequencyHough} pipeline for the search of periodic gravitational wave signals over the full LIGO O2 dataset, with a significant improvement in terms of covered parameter space, and also in terms of results for frequency above 1500 Hz, with respect to previous published results \cite{ref:allsky_o2}. Second, the upper limits have been used to constrain the emission of gravitational waves by boson clouds that have been predicted to spontaneously form around spinning black holes, assuming the emission is nearly monochromatic and dominated by the fundamental scalar mode. Specifically, exclusion regions in the space of black hole/boson masses have been computed for different values of black hole initial spin, boson cloud age and distance. This is the first time results from a real search for gravitational waves have been used to this purpose. 
 We find that with O2 data a range of boson masses, roughly $(1.1\times 10^{-13}-4\times 10^{-13})$ eV for high initial black hole spin, and $(1.2\times 10^{-13}-1.8\times 10^{-13})$ eV for moderate spin,  can be excluded with strong confidence. Qualitatively similar results would be obtained by using upper limits produced by the other \textit{all-sky} search pipelines that run over O2 data, \textit{SkyHough} and \textit{Time Domain F-statistic}  \cite{ref:allsky_o2}, and those obtained over O1 data by the \textit{Einstein@Home} framework (over a much smaller frequency range, 20-100 Hz) \cite{ref:einstein_o1}. The findings discussed in this paper are complementary to those obtained from black hole spin measurements in X-ray binaries, which tend to rule out the mass range from $6\times 10^{-13}$ eV to $10^{-11}$ eV for non-interacting scalar bosons, as a consequence of the measurement of black hole spins as large as 0.98 \cite{ref:cardo2018}. These measures are indeed affected by systematic uncertainties: results can significantly depend on the used method, and are model dependent, as e.g. accretion affects both the mass and the spin of black holes (see \cite{ref:ghosh2019} and \cite{ref:kraw} for more details). Our results are more robust as they refer to isolated black holes and do not rely on electromagnetic observations. 

Although not used in this paper, we have recently developed a robust \textit{all-sky} semi-coherent analysis method which can handle non-monochromatic signals characterized by un-modeled frequency fluctuations, which could be due to some still unpredicted process affecting the gravitational-wave emission. For such signals its sensitivity gain depends on the time scale for the frequency variation and the search setup and, roughly speaking, can range from a few percent to a factor of 3-4, see \cite{ref:danto2018} for more details.
See also \cite{ref:isi2019} for another robust procedure, tailored to directed searches of periodic signals from boson clouds/BH systems with known location.

The chance of detection, or the capability to exclude larger portions of parameter space, will improve analyzing data from the current LIGO-Virgo O3 run and beyond, ultimately helping to shed light on the fascinating connections among particle physics and black holes.    

\begin{acknowledgments}
This research has made use of data obtained from the Gravitational Wave Open Science Center (https://www.gw-openscience.org), a service of LIGO Laboratory, the LIGO Scientific Collaboration and the Virgo Collaboration. LIGO is funded by the U.S. National Science Foundation. Virgo is funded by the French Centre National de Recherche Scientifique (CNRS), the Italian Istituto Nazionale della Fisica Nucleare (INFN) and the Dutch Nikhef, with contributions by Polish and Hungarian institutes.
We thank the LIGO/Virgo Collaboration continuous wave working group for helpful discussions.
We also wish to acknowledge the support of CNAF and Nikhef/SURFSara computing centers where the analysis, from which the upper limits presented in this paper have been derived, has been done.
\end{acknowledgments}





\end{document}